\documentclass[preprint,preprintnumbers,amsmath,amssymb]{revtex4}
\usepackage{graphicx}
\usepackage{dcolumn}
\usepackage{bm}

\begin{document}
\title{Separation of long DNA chains using non-uniform electric field: \\a numerical study}
\author{Shin-ichiro Nagahiro}
 \email{nagahiro@miyagi-ct.ac.jp}
\affiliation{%
Department of Mechanical Engineering, Miyagi National College of Technology, Nodayama, Natori, 981-1239, Japan}%
\author{Satoyuki Kawano}
\affiliation{%
Graduate School of Engineering Science, Osaka University, Osaka, 560-8531, Japan}%
\author{Hidetoshi Kotera}
\affiliation{%
Graduate School of Engineering, Kyoto University, Kyoto 606-8501, Japan}%
\date{\today}
\begin{abstract}
We study migration of DNA molecules through a microchannel with a series of electric traps controlled by an ac electric field. We describe the motion of DNA  based on Brownian dynamics simulations of a beads-spring chain. The DNA chain captured by electric field escapes due to thermal fluctuation. We find that the mobility of the DNA chain would depend on the chain length; the mobility sharply increases when the length of the chain exceeds a critical value, which is strongly affected by the amplitude of the applied ac field. 
Thus we can adjust the length regime, in which the microchannel well separates DNA molecules, without changing structure of the channel. We also present a theoretical insight into the relation between the critical chain length and the strength of binding electric field.  
\end{abstract}
\maketitle

\section{Introduction}
Gel electrophoresis is commonly used in biological science to separate out DNA molecules according to their size. A longer DNA has lower electrophoretic mobility in the random environment of a gel, the separation therefore would be possible using dc electric field.
However, it is well known that gel electrophoresis loses its efficiency for DNA molecules longer than about 20k base pairs (bp), beyond which the length dependence of the mobility gradually disappears. Pulsed-field electrophoresis has been widely used for separation of longer DNA molecules. 
This method also has a limit, both with respect to speed and the size, with an upper limit about 10 Mbp.

On the other hand, separation devices are required to incorporate into a mili-meter size chip with the developments of highly integrated bioanalysis systems --- the so called micro total analysis systems ($\mu$-TAS) or lab-on-a-chip devices \cite{burns, lee}. However introducing gels into microchannels may be a difficult problem. Gel-free separation techniques \cite{ajdari, han1, han2, baka, schmal} therefore have been proposed recently and attracted considerable amount of attention. Ajdari and Prost proposed a gel-free separation technique which uses free-flow electrophoresis together with a series of trapping by induced-dipole forces \cite{ajdari}. They argued from their theoretical analysis that, in their model, the mobility of a DNA chain monotonically decreases with its size.   
More recently, Han and Craighead have been investigated a designed microchannel (entropic trap arrays) which has alternating thin and thick regions \cite{han1, han2}. They found that longer DNA molecules trapped in the thick regions for shorter time. Thus a shorter DNA has lower mobility. This fact has been confirmed theoretically \cite{saka} and numerically \cite{Streek1}.
We note that, as mentioned in Ref. \cite{han1}, the free energy landscape of the above two systems are quite similar, the relation between mobility and DNA size shows the clear contrast. This discrepancy may come from the fact that, to consider escape of DNA chains from a electrode, Ajdari and Prost regarded the chain as a point particle. Contrary, it is experimentally observed that when a DNA molecule escapes from an entropic trap, the deformation of DNA plays an important role \cite{han1}. 

In this paper, we revisit the Ajdari and Prost's model and investigate the mobility of DNA chains along the channel, taking the deformation of DNA molecule into account. Our simulation verifies that longer DNA is trapped for shorter time and the mobility suddenly increases when the chain exceeds a critical length. Furthermore, we found that this critical chain length is strongly affected by the voltage applied across the electrodes. Thus, we could adjust the regime in which the device well separates DNA chains without changing the structure of the channel. These numerical insights will help us to design practical separation devices. 

This paper is organized as follows. In Section II, our separation device and the simulation model are descriced. In Sec. III.A, result of our simulation is discussed, and in Sec. III.B, a scaling analysis is presented and compared with the simulation. In section IV, we summarize our results.
\section{The simulation model}
To our simulation, we adopt a bead-spring model, which represents a DNA molecule as $N$ beads or monomers of diameter $\sigma$. Neighboring beads are connected by harmonic springs with constant $k$ as
\begin{equation}
\frac{U^{\rm sp}(r)}{k_{\rm B}T} = \frac{1}{2}k\left(\frac{r}{\sigma}\right)^2,
\label{eq:spring}
\end{equation}
where $r$ is distance between neighboring beads, $k_{\rm B}$ the Boltzmann constant and $T$ the temperature. 
To take the excluded volume effect into account, all beads interact each other with a Weeks-Chandler-Andersen (WCA) potential 
\begin{eqnarray}
\frac{V^{\rm WCA}(r)}{k_{\rm B}T}&=&
\begin{cases}
	\biggl(\dfrac{\sigma}{r}\biggl)^{12}-\biggl(\dfrac{\sigma}{r}\biggl)^6 +\dfrac{1}{2},&\dfrac{r}{\sigma}<2^{1/6}\\
	~~~0, & {\rm otherwise}.
\end{cases}\label{eq:wca}
\end{eqnarray}
where $r$ is the inter-beads distance.

Let us write the thermal fluctuation force acts on the $i$th bead as $\bm \eta_i(t)$. We suppose that the viscous drag is proportional to the bead velocity with friction constant $\zeta$. A general theorem from statistical mechanics relates the random force and the friction constant as
\begin{equation}
\langle \bm\eta_i(t)\bm\eta_j(t') \rangle = 2k_{\rm B}T\zeta\delta_{ij}\bm I\delta(t-t'),
\end{equation}
where $\bm I$ is the $3\times3$ identity tensor, $\delta_{ij}$ the Kronecker delta and $\delta(t-t')$ the delta function. The random force in our simulation is set so as to satisfy this relation in a discretized manner.

\begin{figure}[t]
\begin{center}
\includegraphics[width=7.0cm, keepaspectratio]{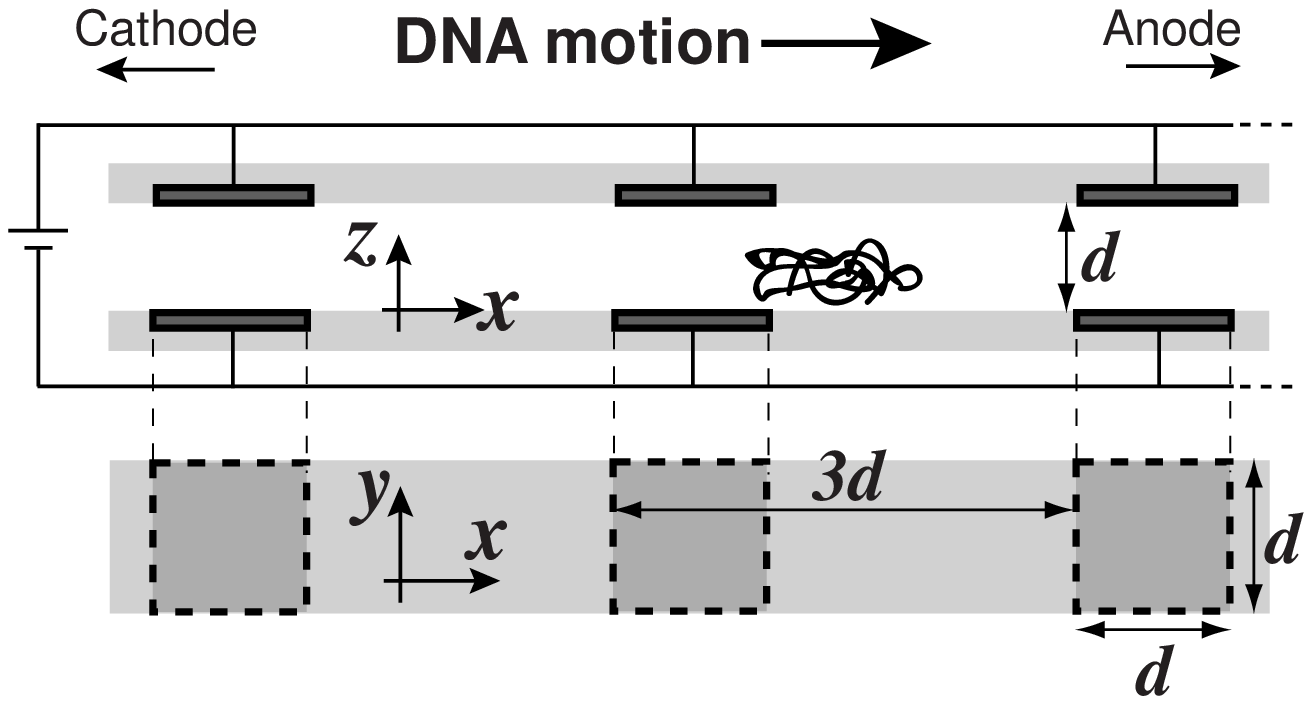}
\caption{A schematic views of a channel with electric traps in $xy$ and $yz$-plane. The electrodes are square in shape and placed along the channel with interval $3d$.}
\label{fig:eta}
\end{center}
\end{figure}

Schematic views in $xz$-plane and $xy$-plane of the channel are shown in Fig. \ref{fig:eta}. We simulate the motion of beads-spring chains inside this channel whose cross-section is a square with side $d$. For $x$ direction, periodic boundary condition with distance $L=9d$ is assumed both for the electric field and the chain motion unless mentioned particularly. We assume that the beads do not adhere to the channel's wall; the WCA potential also applied between the beads and the wall. In the corners, the two repulsive forces come from different directions are summed up. 

Each bead carries a electric charge $q_{\rm net}$. For strongly charged polyelectrolytes such as DNA molecules, the charge $q_{\rm net}$ should be regarded as the renormalized one due to the counterion condensation. 
According to the Oosawa-Maining theory \cite{oosawa}, the charge of DNA, immersed in monovalent salt, is neutralized by $76\%$. Hence we set, for the bead represents $m$ base pairs, $q_{\rm net} \simeq 2.0(1.0-0.76)me^- = 0.48me^-$ ($e^-$ is elementary electric charge).

Next we describe electric trapping of DNA molecules. A charged particle in solution is readily polarized by external electric field due to the cloud of counterions and the charge itself. In the presence of oscillating electric field $\bm E^*$, the polarization energy of this particle could be written as $\Delta U =\alpha|\bm E^*|^2/2$, with $\alpha$ as polarizability. This energy gives force $\bm f^*=-{\rm grad}(\Delta U)$, which drives the particles toward the regions of stronger field amplitude. Some experiments confirmed that the trapping of DNA in the strongest field amplitude could occur over the range of frequency 1Hz $\sim$ 1kHz \cite{charles}. 

\begin{figure}[t]
\begin{center}
\includegraphics[width=7.0cm, keepaspectratio]{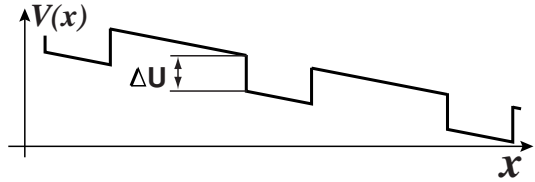}
\caption{A sketch of the electric potential in Ajdari and Prost's channel.}
\label{fig:ratchet}
\end{center}
\end{figure}

Ajdari and Prost theoretically considered the one-dimensional 
motion of a DNA molecule dragged by an static electric field. An ac electric field is applied perpendicular to the direction of DNA migration. In their model, it is assumed that the frequency is high enough and so that the molecule feel a ratchet like potential $V(x)$ as depicted in Fig. \ref{fig:ratchet}. In our simulation, we set ``parallel-plate condensers" along the channel as shown in Fig. \ref{fig:eta} and apply dc voltage across them. We also apply an uniform electric field $\bm E_0$ in $x$-direction, which drives the DNA chain along the channel. The potential $\phi$ inside the channel is obtained by solving the Laplace equation $\nabla^2\phi=0 $ with the boundary element method under the boundary conditions of $\phi=\pm \Delta U$ on the electrodes and $\bm s\cdot\nabla\phi=0$ on the other wall, where $\bm s$ is the surface normal. 
The potential $\phi$ obtained is approximately equivalent to the $V(x)$ in $x$ direction. The electric force acts on a bead is given as ${\bm f}^{\rm el}_i= q_{\rm net}\{ \bm E_0-\nabla\phi(\bm r_i)\}$ where $\bm r_i$ is position of $i$-th bead. Neglecting inertia, we write the equation of motion for $i$-th bead as
\begin{equation}
\zeta\dot{\bm r}_i = \bm f^{\rm int}_i + \bm f^{\rm el}_i + \bm\eta_i,
\label{eq:eqMo}
\end{equation}
where $\bm f^{\rm int}_i$ represents sum of the bead-bead and bead-wall interactions. 

The natural units of our simulation are defined as follows; we use $\sigma$ as the unit length, $k_{\rm B}T$ as the unit energy, ${\cal E}\equiv k_{\rm B}T/\sigma q_{\rm net}$ as the unit electric field strength and $\tau=\zeta\sigma^2/k_{\rm B}T$ as the unit time. 
The channel size $d=20\sigma$. The spring constant $k$ is set to $10^2$, in which condition the equilibrium length of the springs would be $0.85\sigma$ and no chain crossing occurs \cite{Streek2}. We integrated the Eq. (\ref{eq:eqMo}) with second order stochastic Runge-Kutta algorithms \cite{rebecca} using a time step $\Delta t=1.75\times10^{-4}\tau$. The random force is represented as $\bm \eta_i=(k_{\rm B}T/\sigma)\sqrt{6\tau/\Delta t}\bm \psi_i$, where $\bm \psi_i$ is a random vector which has independent components uniformly distributed on $[-1,1]$. We used  Mersenne twister algorithms\cite{makoto} to generate $\bm \psi_i$. Initial conformation of a chain is set by putting beads with interval $\sigma$ on a trajectory obtained from self-avoiding random walk. 

Finally, we mention how the parameters of our model can be related to those of an experiments. We first regard $\sigma=50{\rm nm}$, order of the persistence length of DNA\cite{hage, yoshi}, then one bead corresponds to $167$ base pairs. The electric field ${\cal E}=65$V/cm under condition of $T=300$K. Using the stokes' formula $\zeta=6\pi\sigma\eta$ with $\eta$ as the viscosity of water, we obtain $\tau=5.1\times10^{-4}$sec.

The present model neglects the effect of conterions and hydrodynamic interactions. Adequacy of the simplifications were precisely discussed in Ref.\cite{Streek2}. This model has been successfully described the migration of DNA in gel \cite{Deutsch, Noguchi} and in a channel with entropic traps \cite{Streek1, Streek2}. We hence consider that the model may reproduce experimental results qualitatively and could fit to experimental results with a few adjustable parameters.

\section{Discussion}
\subsection{Results of simulation}
\begin{figure}[tbp]
\begin{center}
\includegraphics[width= 8.0cm, keepaspectratio]{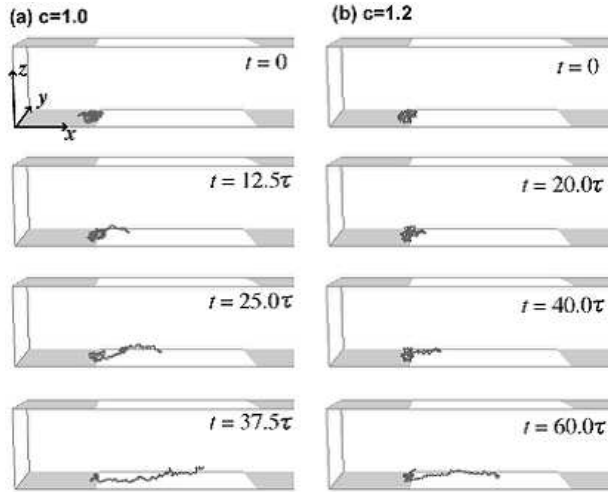}
\caption{A DNA chain of $N=80$ which escapes from a electric trap under the conditions of (a) $c=1.0$  and (b) $c=1.2$. The gray rectangular parts indicate electrodes. The insets in right hand side show a view in $xy$-plane. }
\label{fig:snap}
\end{center}
\end{figure}

\begin{figure}[tbp]
\begin{center}
\includegraphics[width= 6.5cm, keepaspectratio]{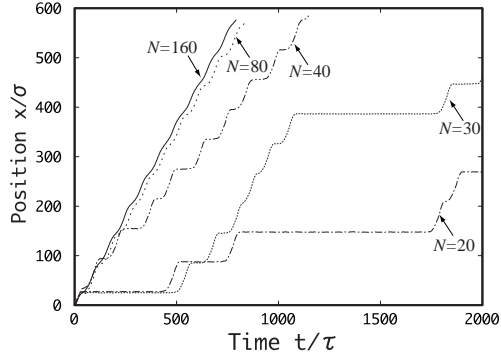}
\caption{Trajectories of DNA chains of $N=$ 20, 30, $\cdots$, 160 $(c=1.0)$. We plot only the $x$-component of the center of mass.}
\label{fig:orbit}
\end{center}
\end{figure}

In this subsection, we show the results of our simulation for DNA migration in the microchannel. Here we set $\bm E_0={\cal E}\hat{\bm x}$, where $\hat{\bm x}$ is the unit vector parallel to the $x$ axis, and describe the applied voltage across each electrode with a dimensionless parameter $c\equiv 2\Delta U/{\cal E}d.$  
The number of beads ranges from $N=5$ to $N=10^3$, which corresponds to DNA molecules from $0.8$ to $170$ kbp. 
We focus on a dimensionless mobility $\tilde{\mu}\equiv\mu{\cal E}\sigma/\tau$ of a single isolated chain as a function of $c$ and $N$.
Figure \ref{fig:snap} shows the snapshots of our simulation, in which a chain of $N=80$ escapes from an electric trap under the conditions of $c=1.0$ and $1.2$. We observe that the chain is stretched during the migration to the next trap. 
Figure.  \ref{fig:orbit} shows trajectories for chains of length $N=20,~30,\cdots, 160$ at $c=1.0$. The flat regions of the trajectories indicate the trapping  of chains on a electrode. Shorter chains tend to be trapped for longer time and longer chains only slightly affected by the traps. To evaluate mobility, we fit the trajectory of which the chain has passed through a trap 10 times, by a linear function. This slope is the dimensionless mobility $\tilde\mu$. 

\begin{figure}[tbp]
\begin{center}
\includegraphics[width= 6.5cm, keepaspectratio]{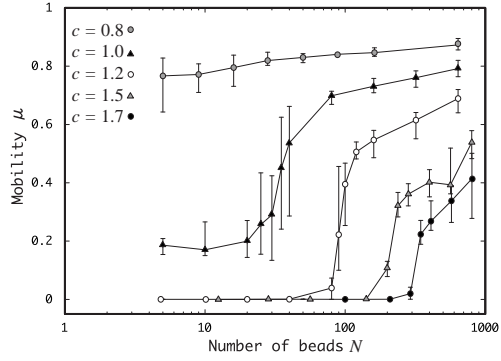}
\caption{Mobility of DNA chains for $c=$ 0.8, 1.0, $\cdots$, 1.7. as the function of $N$. The errorbars show the maximum and minimum values of mobility in 10 traials.}
\label{fig:mobi}
\end{center}
\end{figure}

We show the relation between $\tilde\mu$ and $N$ in Fig. \ref{fig:mobi}. Each plot is the average of 10 trials. For $c=0.8$, the mobility $\tilde \mu$ gently increases and for $c\geq1.0$ a sharp increase of mobility appears. In the case of $c\geq1.2$ short chains less than $N\simeq10^2$ are permanently trapped on a electrode and the resultant mobility would be zero.  In the migration of long chains $N\simeq 800$ under the condition of $c=1.5$ and $1.7$, we observe that the mobility again rises. This is because the chain is stretched two or three times longer than the space between adjacent electrodes and the trapping by electrodes would be less efficient. 
In this regime, to obtain the theoretical understanding in the mobility is not easy. Thus this would be our future work. In the following subsection, we consider the dynamics of DNA chains trapped by only one electrode.

\subsection{Critical chain length $N_{\rm c}$}
The microchannel could be efficient at separating DNA molecules in region where mobility $\tilde\mu$ sharply depends on chain length. Let us write the critical chain length as $N_c$, at which the mobility shows the sudden increase. Figure \ref{fig:mobi} suggests that, by varying $c$, we can ``regurate" the microchannel for the DNA moleucles we try to separate. In the following, we hence consider the relation between $N_{\rm c}$ and $c$.

\begin{figure}[tbp]
\begin{center}
\includegraphics[width= 6.5cm, keepaspectratio]{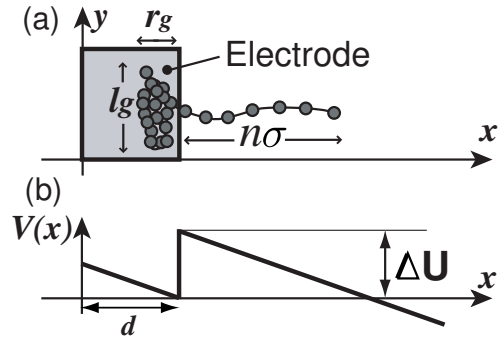}
\caption{(a) A schematic view of a model DNA (bead-spring chain) escaping from a trap. (b) Electric potential energy experienced by a bead.  }
\label{fig:theor_view}
\end{center}
\end{figure}

\begin{figure}[tbp]
\begin{center}
\includegraphics[width=6.0cm, keepaspectratio]{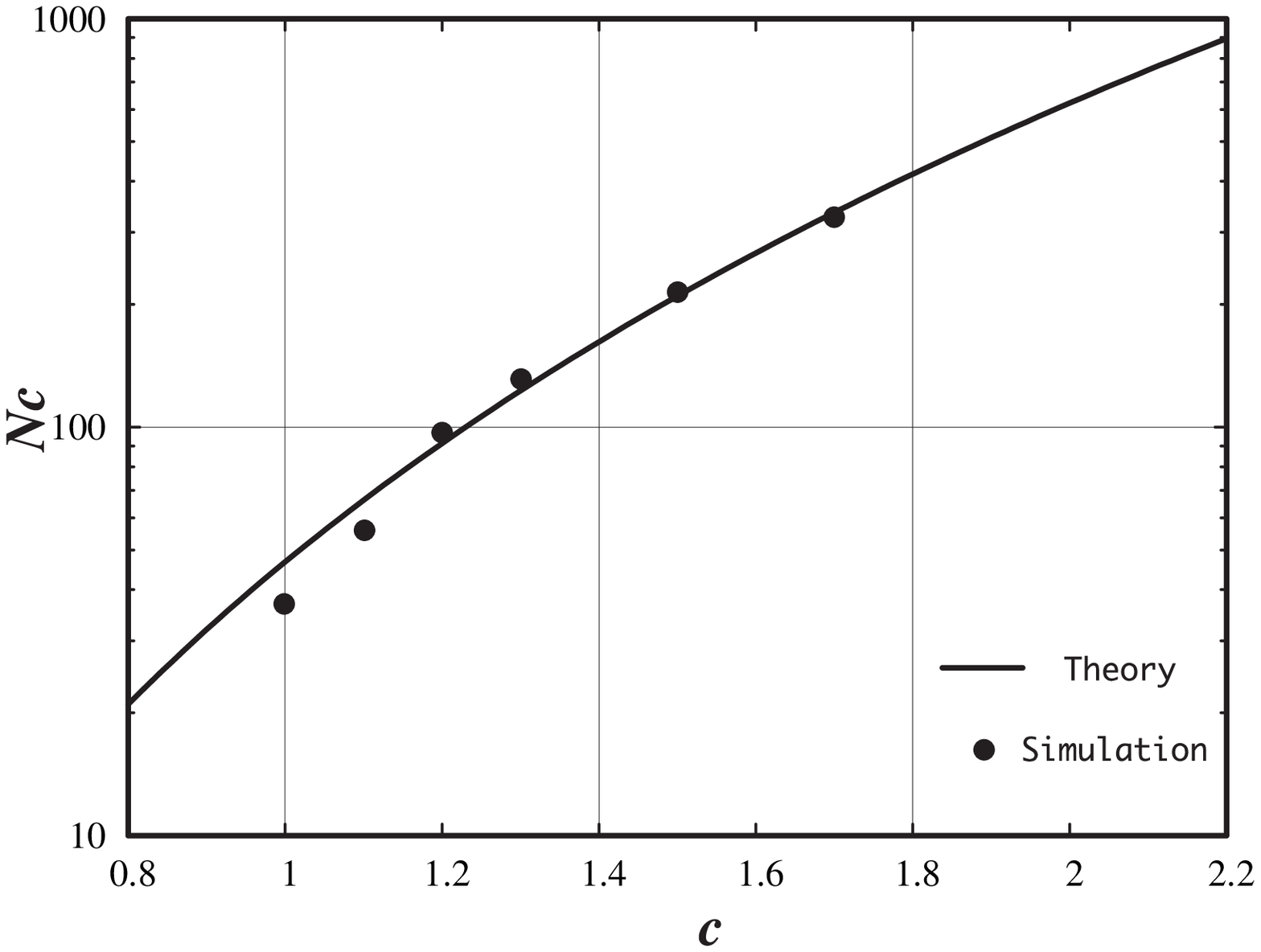}
\caption{The black circles represent the critical chain length estimated from our simulation. The solid line is a plot of Eq. (\ref{eq:theoN}) fitted to our simulation with parameter $A=3.4$. }
\label{fig:n_c}
\end{center}
\end{figure}

Let us write a trapping lifetime of DNA chains as $\tau_t(c, N)$. The dimensionless mobility could be written as \cite{han1}
\begin{equation}
\tilde{\mu}= \frac{t_0}{t_0+\tau_t},
\end{equation}
where $t_0$ is the transit time between two adjacent electrodes. Figure \ref{fig:theor_view}(a) is a schematic view of a bead-spring chain trapped on a electrode, in which $n<N$ beads of the chain have overcome a electric barrier $\Delta U$. Each bead experiences the ratchet like potential $V(x)$ as shown in Fig. \ref{fig:theor_view}(b), where we set $V(x)|_{x=d}=0$.  To calculate trapping life time $\tau_t$, we roughly estimate total free energy difference $\Delta F(n)$ due to the escape of $n$ beads. Let $r_g$ and $l_g$ be the typical sizes of a ``blob" trapped on a electrodes. These two could be given by \cite{gennes} 
\begin{eqnarray}
r_g &\sim&\sigma(N-n)^{\nu_r},\label{eq:rg}\\
l_g &\sim& \sigma(N-n)^{\nu_l}.\label{eq:lg}
\end{eqnarray}
Unfortunately, we cannot analytically calculate the exponents $\nu_r$ and $\nu_l$. However, if we neglect the side walls at $y=0$ and $d$, our simulation evaluates $\nu_r\simeq0.44$ and $\nu_l\simeq0.37$. 
Using the lengthes $r_{g}$ and $l_{g}$, we can roughly estimate the electric potential energy of a blob trapped on a electrode $U_{\rm blob}$ as
\begin{equation}
U_{\rm blob} =Aq_{\rm net}\frac{l_gr_g^2{\cal E}}{\sigma^2},
\end{equation}
where $A$ is constant. When $n$ beads have overcome a barrier $\Delta U$, the whole chain gains potential energy $nq_{\rm net}\Delta U$ and loses $\sigma q_{\rm net}{\cal E}n^2/2$ as the $n$ beads go down the slope. The potential energy  of the chain therefore could be written as
\begin{eqnarray}
U(n)&\simeq& Aq_{\rm net}\frac{l_gr_g^2{\cal E}}{\sigma^2}+nq_{\rm net}\Delta U-\sigma q_{\rm net}{\cal E}\frac{n^2}{2}\nonumber,\\
&=& q_{\rm net}\sigma{\cal E} \left\{A(N-n)^{\gamma}+\frac{cd}{2\sigma}n-\frac{n^2}{2}\right\},
\end{eqnarray}
where $\gamma\equiv 2\nu_r+\nu_l\simeq 1.25$. Because the increase of entropic free energy is proportional to $nk_{\rm B}T$, the total free energy difference $\Delta F(n)$ is given by 
\begin{eqnarray}
\frac{\Delta F(n)}{k_{\rm B}T}&\sim&\frac{U(n)-U(0)}{k_{\rm B}T}+n\nonumber\\
&=&A\left\{\left(N-n\right)^{\gamma}-N^\gamma\right\}+\frac{cd}{2\sigma}n-\frac{n^2}{2}+n.\label{eq:freeEng}
\end{eqnarray}
In Fig. \ref{fig:theor_view}(b), we plot $\Delta F(n)$ as a function of ratio $n/N$. 
For $c>1$ the $\Delta F(n)$ increases at a small $n$ and reaches a peak at $n=n^*$, while for $c=0.5$ it monotonically decreases with $n$.
Let $\Delta F_{\rm max}$ be the maximum of $\Delta F(n)$. 
Trapping occurs when $d(\Delta F)/dn|_{n=0}>0$, and $\tau_t$ is given by
\begin{equation}
\tau_t\sim\exp(\Delta F_{\rm max}/k_{\rm B}T).
\end{equation}
We can calculate $\Delta F_{\rm max}$ by solving the following equation
\begin{eqnarray}
\left.\frac{d}{dn}\frac{\Delta F(n)}{k_{\rm B}T}\right|_{n=n^*}=\hspace{11em}\nonumber\\
-A\gamma N^{\gamma-1}\left(1-\frac{n^*}{N}\right)^{\gamma-1}+\frac{cd}{2\sigma}-n^*+1=0.
\label{eq:maxEq}
\end{eqnarray}
Here we assume that number of beads $N$ is large enough, and neglect 
the order $N^{\gamma-2}$ and the higher terms. Then we obtain
\begin{equation}
n^*\simeq\frac{cd}{2\sigma}+1-A\gamma N^{\gamma-1}.
\end{equation}
The sharp increase of $\tilde{\mu}$ appears when the free energy barrier $\Delta F_{\rm max}=0$. From Fig.\ref{fig:theor_view}(b), we know that this condition occurs only when $n^*=0$. We therefore obtain
\begin{equation}
N_c \simeq \left(\frac{cd/2+\sigma}{\sigma\gamma A}\right)^{\frac{1}{\gamma-1}},
~~~\frac{1}{\gamma-1}=4.0
\label{eq:theoN}
\end{equation}
with $A$ as a fitting parameter. To estimate the value of $N_c$ from simulation data, we adopted the following procedure: first, we obtained the maximum and minimum mobility ($\mu_{\rm max},~\mu_{\rm min}$) for a fixed $c$ in region $5<N<800$. We write number of beads ${N_c}'$ with this length the simulation gives mobility $(\mu_{\rm max}+\mu_{\rm min})/2$. Figure \ref{fig:n_c} shows that Eq. (\ref{eq:theoN}) is in good agreement with ${N_c}'$ with one adjustable parameter $A=3.4$. 
\section{concluding remarks}
We have presented Brownian dynamics simulations of DNA migration in channels with electric trappings. The chain length dependence of mobility observed in our simulation does not agree with the theory presented by Ajdari and Prost \cite{ajdari}. We conclude that this discrepancy is due to their simplification, in which DNA molecules escapes from electric traps as a point particle, because our numerical simulations demonstrated that the deformation of DNA chain plays an important role when it escapes from the traps. 

Furthermore, the critical chain length $N_c$ is strongly affected by the voltage applied across the electrodes. From the rough consideration, we found a simple scaling relation between $N_c$ and the parameters of the present system, which agrees with simulations with one adjustable parameters.   

\begin{acknowledgments}
We wish to acknowledge helpful discussions with Takahiro Sakaue and Youhei Maruyama. This study is a part of joint researches, which are aiming at developing the basis of technology COE in nano-medicine, carried out through Kyoto City Collaboration of Regional Entities for Advancing Technology Excellence (CREATE) assigned by Japan Science and technology Agency (JST).
\end{acknowledgments}

\begin{figure}[tbp]
\begin{center}
\includegraphics[width= 6.5cm, keepaspectratio]{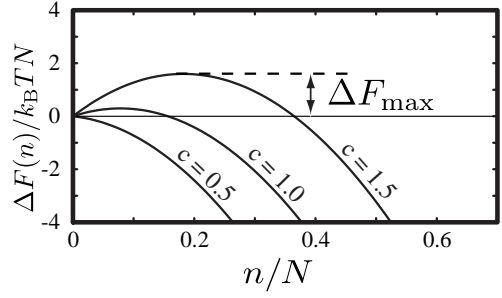}
\caption{The plot of Eq. (\ref{eq:freeEng}) for $c=0.5$, $1.0$ and $1.5$ ($A=3.4$).  }
\label{fig:freeEng}
\end{center}
\end{figure}

\begin{thebibliography}{99}
\bibitem{burns} M. A. Burns et al., Science {\bf 282}. 484 (1998)
\bibitem{lee}S. J. Lee and S. Y. Lee, Appl. Microbiol. Biotechnol. {\bf 64}, 289 (2004)
\bibitem{ajdari} A. Ajdari and J. Prost, Proc. Natl. Acad. Sci. USA {\bf 88}, 4468 (1991)
\bibitem{han1} J. Han, S. W. Turner and H. G. Craighead, Phys. Rev. Lett. {\bf 83}, 1688 (1999)
\bibitem{han2} J. Han and H. G. Craighead, Science {\bf 288}, 1026 (2000), J. Han and H. G. Craighead, Anal. Chem. {\bf 74}, 394 (2002)
\bibitem{baka} O. Bakajin, T. A. J. Duke, J. Tegenfeldt, C.-F. Chou, S. S. Chan, R. H. Austin, and E. C. Cox, Anal. Chem. {\bf 73}, 6053 (2001)
\bibitem{schmal} D. Schmalzing, L. Koutny, A. Adourian, P. Belgraderdagger, P. Matsudaira, and D. Ehrlich, Proc. Natl. Acad. Sci. USA {\bf 94}, 10273  (1997)
\bibitem{saka} T. Sakaue, Eur. Phys. J. E (2006)
\bibitem{Streek1} M. Streek, F. Schmid, T. T. Duong and A. Ros, J. Biotechnol. {\bf 112}, 79 (2004)
\bibitem{Deutsch} J. M. Deutsch, Phys. Rev. Lett. {\bf 59}, 1255 (1987)
\bibitem{Noguchi} H. Noguchi and M. Takasu, J. Chem. Phys. {\bf 114}, 7260 (2001)
\bibitem{Streek2} M. Streek, F. Schmid, T. T. Duong, D. Anselmetti and A. Ros, Phys. Rev. E {\bf 71}, 011905 (2005)
\bibitem{charles} C. L. Asbury and G. van den Engh, Biophys. J. {\bf 74}, 1024 (1998)
\bibitem{oosawa} F. Oosawa, J. Polymer Sci. {\bf 13}, 421 (1957), G. S. Manning, Q. Rev. Biophys. {\bf 11}, 179 (1978)
\bibitem{hage} P. J. Hagerman, Annu. Rev. Biophys. Biophys. Chem. {\bf 17}, 265 (1988)
\bibitem{yoshi} K. Yoshikawa, Y, Matsuzawa, K. Minagawa, M. Doi and M. Matsumoto, Biochem. Biophys. Res. Commun. {\bf 188}, 1274 (1992)
\bibitem{rebecca} R. L. Honeycutt, Phys. Rev. A {\bf 45}, 600 (1992)
\bibitem{makoto} M. Matsumoto and Y. Nishimura, ACM Trans. Modeling Comp. Simul. {\bf 8}, 3 (1998)
\bibitem{gennes} P. -G. de Gennes, {\it Scaling Concepts in Polymer Physics} (Cornell University Press, Ithaca, NY, 1979), T. Sakaue and E. Rapha{\"{e}}l, Macromolecules {\bf 39}, 2621 (2006)
\end{thebibliography}
\end{document}